\documentclass[epjCONF]{svjour}
\usepackage{graphics}
\usepackage[varg]{txfonts} 
\usepackage[latin1]{inputenc}
\session-title{40th LIAC}
\begin{document}
\title{Hot Subdwarf Formation: Confronting Theory with Observation}
\author{S. Geier\inst{1}\fnmsep\thanks{\email{geier@sternwarte.uni-erlangen.de}}}
\institute{Dr.\,Karl Remeis-Observatory \& ECAP, Astronomical Institute, Friedrich-Alexander University Erlangen-Nuremberg, Sternwartstr.~7, 96049 Bamberg, Germany}
\abstract{The formation of hot subdwarf stars is still unclear. Both single-star and binary scenarios have been proposed to explain the properties of these evolved stars situated at the extreme blue end of the horizontal branch. The observational evidence gathered in the last decade, which revealed high fractions of binaries, shifted the focus from the single-star to the binary formation scenarios. Common envelope ejection, stable Roche lobe overflow and the merger of helium white dwarfs seemed to be sufficient to explain the formation of both the binary as well as the remaining single hot subdwarfs. However, most recent and rather unexpected observations challenge the standard binary evolution scenarios.
} 
\maketitle
\section{Introduction}\label{intro}

Subluminous stars are less luminous than main-sequence stars of similar spectral type and are therefore situated below the main sequence in the Hertzsprung-Russell-Diagram (HRD). Since main sequence stars are also called dwarfs in contrast to the giant stars of the same spectral type, subluminous stars were therefore called subdwarfs (see review by Heber \cite{heber09}). 

Depending on their spectral appearance, hot subdwarf stars can be divided into subclasses. The original definition of hot subdwarf B stars comes from Sargent \& Searle \cite{sargent68}, who introduced it for stars with colours similar to main sequence B stars, but with much broader Balmer lines. Hot subdwarf B (sdB) stars show strong and broad Balmer and weak (or no) He\,{\sc i} lines. sdOB stars show strong and broad Balmer lines as well as weak lines from He\,{\sc i} and He\,{\sc ii}, while sdO stars only display weak He\,{\sc ii} lines besides their strong Balmer lines. He-sdBs are dominated by strong He\,{\sc i} and sometimes weaker He\,{\sc ii} lines. Hydrogen absorption lines are shallow or not present at all. He-sdOs show strong He\,{\sc ii} and sometimes weak He\,{\sc i} lines. Balmer lines are not present or heavily blended by the strong He\,{\sc ii} lines of the Pickering series \cite{moehler90}. 

In the HRD the hot subdwarf stars are situated at the blueward extension of the Horizontal Branch (HB), the so called Extreme or Extended Horizontal Branch (EHB) \cite{heber86,saffer94}. The most common class of hot subdwarfs, the sdB stars, are located on the EHB and are therefore considered to be core helium-burning stars. They have very thin hydrogen dominated atmospheres ($n_{\rm He}/n_{\rm H}\leq0.01$) and masses around $0.5\,M_{\rm \odot}$ \cite{heber86}. Their effective temperatures ($T_{\rm eff}$) range from $20\,000\,{\rm K}$ to $40\,000\,{\rm K}$ and their surface gravities ($\log{g}$) are one to two orders of magnitude higher than in main sequence stars of the same spectral type (usually between $5.0$ and $6.0$). They consist of a helium-burning core surrounded by a thin hydrogen-rich envelope ($M_{\rm env}<0.02\,M_ {\rm \odot}$). Also post-AGB objects in a certain evolutionary stage and more massive He-stars, like the so called low gravity or luminous He-sdOs \cite{jeffery08} belong to the class of hot subdwarfs and are situated between the EHB and the main sequence in the HRD. 

The formation of these objects is still puzzling. SdB stars can only be formed, if the progenitor loses its envelope almost entirely after passing the red giant branch (RGB) and the remaining hydrogen-rich envelope has not retained enough mass to sustain a hydrogen-burning shell, which is the case in cooler HB stars. Therefore the star can not evolve in the canonical way and ascend the Asymptotic Giant Branch (AGB). In contrast to this the star remains on the EHB until core helium-burning stops, and after a short time of shell helium-burning eventually reaches the white dwarf (WD) cooling tracks. According to evolutionary calculations the average lifetime on the EHB is of the order of $10^{8}\,{\rm yr}$ \cite{dorman93}. In this canonical scenario the hotter ($T_{\rm eff}=40\,000-80\,000\,{\rm K}$) and much less numerous hydrogen rich sdOs can be explained as rather short-lived shell helium-burning stars evolving away from the EHB.

Some hot subluminous stars are not connected to EHB-evolution at all. Objects with spectra and atmospheric parameters similar to normal sdBs are known, which are situated below the EHB \cite{heber03,otoole06,vennes11,silvotti12}. These objects are considered to be direct progenitors of helium white dwarfs, which descend from the red giant branch. For these post-RGB objects, which cross the EHB, evolutionary tracks indicate masses of about $0.20-0.33\,M_{\rm \odot}$ \cite{driebe98}. In order to form such objects, the mass loss at the RGB has to be more extreme than in the case of EHB stars. Objects of even lower masses have been discovered recently and are known as extremely low-mass (ELM) WDs \cite{brown10,maxted11,kilic12}. 

\subsection{Single star formation and evolution scenarios}

The reason for the very high mass loss at the tip of the RGB after the helium flash is still unclear. Several single star scenarios have been proposed: Stellar wind mass loss at the RGB \cite{dcruz96}, helium mixing by internal rotation in the RGB phase \cite{sweigart97} and envelope stripping processes in dense clusters \cite{demarchi96} or through a supernova explosion of the companion in a binary system \cite{marietta00}. All these scenarios require either a fine-tuning of parameters or extreme environmental conditions which are unlikely to be met for the bulk of the observed subdwarfs in the field. Recent ideas invoke tidally induced stellar winds or the ejection of the envelope due to positive binding energy \cite{han12}. Those scenarios may be more generally applicable. 

The formation of He-sdO/Bs is even more enigmatic. Most (but not all) He-sdOs are concentrated at a very small region in the HRD, slightly blueward of the EHB at $T_{\rm eff}=40\,000-80\,000\,{\rm K}$ and $\log{g}=5.60-6.10$ \cite{stroer07}. The He-sdBs are scattered above the EHB. The late hot flasher scenario provides a possible channel to form these objects \cite{lanz04,miller08}. After ejecting most of its envelope at the tip of the RGB, the stellar remnant evolves directly towards the WD cooling tracks and experiences a late core helium flash there. Helium is mixed into the atmosphere and the star ends up close to the helium main sequence. Depending on the depth of the mixing, stars with more or less helium in the atmospheres and different atmospheric parameters can be formed in this way. However, the required loss of the hydrogen envelope on the RGB still remains unexplained.

\subsection{Hot subdwarf binaries and binary formation scenarios}

The first subdwarf binary with a main sequence F companion visible in the spectrum was discovered by Wallerstein \& Spinrad \cite{wallerstein60}. 
Since then numerous studies searched for such composite systems using photometry from the UV to the infrared \cite{stark03,girven12} as well as spectroscopy \cite{lisker05}. The derived number fraction of sdO/B binaries with main sequence companions (usually K- to F-type) ranges from $\simeq30$ to $40\,\%$ \cite{heber09}. 

Systematic surveys for radial velocity (RV) variable stars revealed that a large fraction of the sdB stars ($40-70\,\%$) are members of close binaries with orbital periods ranging from $\simeq0.07\,{\rm d}$ to $\simeq30\,{\rm d}$ \cite{maxted01,morales03,geier11a,copperwheat11}. Most of the known companions of sdBs in radial velocity variable close binary systems are white dwarfs or late type main sequence stars. Candidates with more massive compact companions like neutron stars or even black holes have been found as well \cite{geier10}.

In contrast to that, the population of He-sdOs observed so far seems to consist mostly of single stars. Only one RV-variable He-sdO has been found in the SPY sample, which corresponds to a fraction of only $3\,\%$ \cite{napiwotzki08}. However, higher fractions have been reported for the He-sdO populations in both the PG \cite{green08} and the MUCHFUSS samples \cite{geier11b}. Since no orbital solutions could be derived so far, these results remain unclear. 

Binary interaction obviously plays an important role in the process of formation and evolution of hot subdwarf stars \cite{mengel76}. 
For the close binary systems common envelope (CE) ejection is the only viable formation channel \cite{paczynski76}. In this scenario two main sequence stars of different masses evolve in a binary system. The more massive one will be the first to reach the red-giant phase and fill its Roche lobe. If the mass transfer to the companion is dynamically unstable, a common envelope is formed. The two stellar cores lose orbital energy, which is deposited within the envelope and leads to a shortage of the binary period. Eventually the common envelope is ejected and a close binary system is formed. It contains a core helium-burning sdB and a main sequence companion. If this companion reaches the red giant branch, another mass transfer phase is possible and can lead to a close binary with a white dwarf companion and an sdB. 

If the mass transfer to the companion is dynamically stable, the companion fills its Roche lobe and the primary slowly accretes matter from the secondary. The companion eventually loses most of its envelope and becomes an sdB. This leads to sdB binaries with main sequence companions and much larger separations. Stable RLOF channel and CE ejection may also be mixed. After a phase of stable RLOF producing a wide binary consisting of a WD and an MS star, a CE ejection phase can lead to the formation of close sdB+WD binary.

An alternative way of forming a single sdB is the merger of two helium white dwarfs \cite{webbink84,iben84}. Close He-WD binaries are formed as a result of two CE-phases. Loss of angular momentum through emission of gravitational radiation will cause the system to shrink. Given the initial separation is short enough, the two white dwarfs eventually merge and if the mass of the merger is high enough, core helium-burning is ignited and a hot subdwarf is formed. 

These three binary channels for sdB formation were adressed in detail by Han et al. \cite{han02,han03}, who performed binary population synthesis studies. The observed distribution of orbital parameters could be reproduced by these simulations at least for the close binary systems. One important result of this study is, that although most sdBs are still expected to have masses around $0.47\,M_{\rm \odot}$ as predicted by single star evolution, the possible mass range is broader. The mass of subdwarfs originating from main sequence stars massive enough to ignite core-helium burning under non-degenerate conditions can be as low as $\simeq0.3\,M_{\rm \odot}$ whereas subdwarfs formed through the WD merger channel may be as massive as $\simeq0.7\,M_{\rm \odot}$. 

\section{Observational challenges}

The observational evidence gathered in the last decade, which revealed a high fraction of both double-lined binary systems with MS companions and short-period, RV-variable binaries, shifted the focus from the single star to the binary formation scenarios elaborated by Han et al. \cite{han02,han03}. The three channels CE-ejection, stable RLOF and WD merger seemed to be sufficient to explain the formation of both binary types as well as the remaining single hot subdwarfs. Furthermore, related issues like the UV-excess in evolved galaxies \cite{han07} and the significantly lower fraction of close binary sdBs observed in globular clusters \cite{monibidin09,monibidin11,han08} could be explained in this context.

However, most recent and rather unexpected observational results challenge the standard binary evolution scenarios. In the following sections, these results will be discussed in detail.

\subsection{Substellar companions}

It has been proposed, that planets and brown dwarfs could also be responsible for the huge loss of envelope mass in the red-giant phase necessary to form hot subdwarfs \cite{soker98}. As soon as the host star evolves into a red giant, close substellar companions (like hot Jupiters) will be engulfed by the stellar envelope. However, whether those objects are able to eject the envelope and survive, evaporate or merge with the stellar core was unclear. Planets around pulsating and close binary sdBs have indeed been discovered \cite{silvotti07,beuermann12}, but these are too far away from their host stars to have interacted in the past. 

In contrast to that, the eclipsing, short-period sdB binary J0820+0008, discovered in the course of the MUCHFUSS project, turned out to have a brown dwarf companion \cite{geier11c}. Additionally, at least one more similar system and several candidates have been discovered, which proves that substellar companions are able to eject a common envelope. Furthermore, the number fraction of substellar companions seems to be comparable to the one of low-mass main sequence stars ($\simeq0.1\,M_{\rm \odot}$) \cite{geier12a}. This means that the substellar channel forms a significant fraction of the sdB binaries and may also be responsible for the formation of single sdBs, because such companions do not necessarily have to survive the CE phase \cite{soker98}. Recently, earth-sized objects in close orbits around a pulsating sdB observed by the Kepler mission have been discovered. Those objects may be the remnants of more massive planets destroyed during the common envelope phase \cite{charpinet11}.

The involvement of substellar companions challenges the standard CE-ejection scenario, because the amount of energy and angular momentum transfered to the envelope during the CE-phase scales with the mass of the engulfed companion. However, in the classical prescription of the CE-process \cite{han02}, the minimum mass for CE-ejection is in the stellar regime. Even if additional internal energy is taken into account, substellar companion should hardly be able to eject the common envelope and to help form sdBs \cite{han12}. 

\subsection{Rotational properties of single sdBs}

The projected rotational velocities of a large sample consisting of single and wide binary sdBs have recently been determined by measuring the broadening of metal lines \cite{geier12b}. All stars in this sample turned out to be slow rotators (${v_{\rm rot}\sin{i}}<10\,{\rm km\,s^{-1}}$). Furthermore, the $v_{\rm rot}\sin{i}$-distributions of single sdBs are similar to those of hot subdwarfs in wide binaries with main-sequence companions as well as close binary systems with unseen companions and periods long enough that tidal effects become negligible. Furthermore, hot blue horizontal branch stars and extreme horizontal branch stars are related in terms of surface rotation and angular momentum. 

The ${v_{\rm rot}\sin\,i}$-distribution of the single sdB stars is particularly hard to understand in the context of the WD merger scenario, because merger remnants are likely to spin fast. Angular momentum may be lost through stellar winds and magnetic fields or an interaction with the accretion disc during the merger. However, even if the merged remnant of two He-WDs should be significantly slowed down, it is unlikely that the merged products would have a ${v_{\rm rot}\sin{i}}$-distribution almost identical to sdBs formed via CE-ejection or maybe stable RLOF. It therefore seems as if the merger channel does not contribute significantly to the observed population of single hydrogen-rich sdO/Bs in contrast to the standard models \cite{han02,han03}. 

\subsection{Empirical mass distribution}

The conclusion drawn from the ${v_{\rm rot}\sin\,i}$-distribution of the single sdB stars is consistent with a study of the empirical mass distribution of sdB stars derived from eclipsing binary systems and asteroseismic analyses \cite{fontaine12}. No sdB stars more massive than $\simeq0.5\,M_{\rm \odot}$ have been clearly discovered yet. Since a significant fraction of such massive sdBs should have been formed by WD mergers, the conclusion is that at least for the hydrogen-rich sdBs mergers are less frequent than predicted by theory.

\subsection{Orbital parameters of double-lined sdB+MS binaries}

The existence of double-lined sdB binaries with main sequence companions has been conveniently explained by the stable RLOF formation channel. Since the progenitor systems of those binaries must have transfered mass in the past, the separation of the formed binaries should not be too large. The standard models predict orbital periods of less than $500\,{\rm d}$ \cite{han02,han03}. 

First evidence that the distribution of separations deviates from the one of main sequence binaries was found from HST photometry with high angular resolution. Since none of the the sdB+MS systems could be resolved, it was concluded that the separations of those binaries must be significantly smaller than the ones of non-interacting main sequence binaries  indicating an interaction in the past evolution \cite{heber02}.

However, most recently the first orbital solutions of double-lined sdB+MS systems could be derived \cite{deca12,oestensen12,barlow12}. Contrary to the predictions from binary evolution theory, all those systems have orbital periods ranging from $500-1200\,{\rm d}$. Furthermore, one system (PG\,1338$+$061, $P=937\,{\rm d}$) was found to have an eccentric orbit ($e=0.15\pm0.02$). Since stable RLOF should be very efficient in circularising the orbit, this discovery indicates that no such mass-transfer has taken place and the sdB must have been formed in a different way. There are hints that some of the studied systems are hierarchical triple systems, where a close binary sdB is orbited by a main sequence companion in a wide orbit \cite{heber02,barlow12}. 

\section{Conclusion}

Although the standard binary evolution scenarios have been very successful in explaining the general properties of the known hot subdwarf population, most recent observations challenge certain aspects of this picture. The CE-scenario, although not well understood in detail, is the only known way to form close binary sdBs. The existence of substellar companions in such systems provides crucial constraints for our understanding of this process. A mechanism is needed to lower the minimum companion mass necessary for CE-ejection. Current ideas include a spin-up of the red-giant envelope before entering the CE-phase \cite{bear10} or a dependence of the CE-efficiency on the mass of the companion \cite{demarco11}.

It is unlikely that He-WD mergers form the bulk of the single hydrogen-rich sdBs. Both their rotational properties and their mass distribution are hardly consistent with this scenario. He-sdOs on the other hand might be possible candidates for a merger origin, since their projected rotational velocities are two to three times higher than the ones of sdBs \cite{hirsch09}. Although this seems not particularly high, it is still significantly different. In addition to the He-WD merger, the merger of an sdB and an He-WD has been proposed as possible formation scenario for He-sdOs \cite{justham10}. Furthermore, the recent discovery of a single, fast-rotating sdB, which may have been formed by a merger event during the CE-phase \cite{geier11d,politano08}, shows that mergers might still play a role in hot subdwarf formation, but maybe in a different way than thought before.

The problem that sdB+MS binaries have longer orbital periods than predicted by the stable RLOF-channel may simply be overcome by using more sophisticated binary evolution models (see discussion in \cite{clausen12} and Han et al. these proceedings). However, the eccentric orbit observed in at least one of these objects seems to be at odds with the whole idea of stable mass-transfer. Recently, the merger of an He-WD and a low-mass MS star was proposed as possible formation scenario. In contrast to the standard merger scenarios, the He-core is swallowed by the MS star thereby creating a more evolved star, that eventually loses its envelope and forms an sdB. This scenario is interesting because it explains the slow rotation of single sdBs in a natural way. Angular momentum is lost via the ejection of the envelope. Should such a merger happen in a triple system, it will form an sdB+MS binary. Since these stars did not interact in the past, eccentric orbits are possible \cite{clausen11}.

%
%

\end{document}